# Optical Properties and Electronic Structures of Intrinsic Gapped Metals: Inverse Materials Design Principles for Transparent Conductors

Muhammad Rizwan Khan[1], Harshan Reddy Gopidi[1], and Oleksandr I. Malyi[1,#]

[1]Centre of Excellence ENSEMBLE³ Sp. z o. o., Wolczynska Str. 133, 01-919, Warsaw, Poland

[#]**Email:** oleksandrmalyi@gmail.com

**Abstract:** Traditional solid-state physics has long correlated the optical properties of materials with their electronic structures. However, recent discoveries of intrinsic gapped metals have challenged this classical view. Gapped metals possess electronic properties distinct from both metals and insulators, with a large concentration of free carriers without any intentional doping and an internal band gap. This unique electronic structure makes gapped metals potentially superior to materials designed by intentional doping of the wide band gap insulators. Despite their promising applications, such as transparent conductors, designing gapped metals for specific purposes remains challenging due to the lack of understanding of the correlation between their electronic band structures and optical properties. This study focuses on representative examples of gapped metals and demonstrates the cases of (i) gapped metals (e.g., $CaN_2$) with strong intraband absorption in the visible range, (ii) gapped metals (e.g., $SrNbO_3$) with strong interband absorption in the visible range, (iii) gapped metals (e.g., $Sr_5Nb_5O_{17}$) that are potential transparent conductors. We explore the complexity of identifying potential gapped metals for transparent conductors and propose inverse materials design principles for discovering new-generation transparent conductors.

In recent years, there has been a growing interest in understanding and manipulating the optical properties of materials for various technological applications, ranging from optoelectronics and photovoltaics to sensing and quantum computing[1-3]. Traditional solid-state physics books[4, 5] teach us that the optical properties of a material are defined by its electronic structure. For instance, wide band gap insulators are transparent due to the large energy band gaps (over 3.2 eV) between their principal valence and conduction bands, which prevents visible light from exciting electrons from occupied to unoccupied states[6-8]. In contrast, metals are opaque due to the continuous nature of their electronic structure, resulting in strong band-to-band transitions and free carrier absorption[9, 10]. This correlation between electronic structure and optical properties allows us to use the color of a compound as an indicator of its conductivity. For example, in semiconductors, the band gap is smaller than in insulators, allowing for the absorption of photons with shorter wavelengths, resulting in a colored sample[11, 12].

Recent studies[13-20] have challenged this classical picture by revealing the existence of intrinsic gapped metals that posse electronic properties distinct from both metals and insulators. Specifically, these materials have a Fermi level in the conduction or valence band, resulting in a large concentration of free carriers without intentional doping but with an internal band gap between their principal band edges. This unique electronic structure makes gapped metals potentially superior compared to materials designed through intentional doping of wide band gap insulators, as gapped metals are not subject to the limitation of doping bottleneck (i.e., Fermi level pinning or defect clustering, restricting the ability to modify material's electronic properties through intentional doping)[13, 21]. In addition, these material features make gapped metals potentially useful for various applications. For instance, Irvine et al.[22] discovered that metallic $SrNbO_3$ could be used for photocatalysis, Hosono et al.[23] reported the first solid-state electride gapped metal, while Zunger et al.[14] demonstrated that intrinsic gapped metals could be transparent conductors.

However, it is important to note that designing gapped metals for specific applications, such as transparent conductors, is not a straightforward task due to the lack of understanding regarding the correlation between their electronic band structures and optical properties. In particular, the criteria for selecting the appropriate atomic identities, composition, and structure to achieve the desired optoelectronic functionality is often uncertain. This uncertainty arises from the complex interplay of free carrier absorption and interband transition, which together define the optical properties of the gapped metals, making a comprehensive understanding of the essential step of the material design. To address this, we focus on representative examples of gapped metals and identify three principal types of intrinsic gapped metals based on their electronic structures (see discussion below). These classifications help shed light on the underlying mechanisms governing the optical properties of these materials, which in turn help to establish the design principles for potential transparent conductors and provide a systematic approach to the discovery of new materials, taking into account the complex interplay of factors that determine the properties of gapped metals.

The optical properties of gapped metals are defined by the superposition of interband transition and the Drude contribution. The latter is characterized by the ratio of carrier concentration ($n$) and effective mass ($m^*$) defining plasma frequencies (i.e., $\omega_p \sim \sqrt{n/m^*}$). When the ratio is large, free carrier absorption makes the material opaque, even if it has a large internal band gap. This behavior is exemplified by the experimentally synthesized $CaN_2$ compound (space group (SG: 139), which is known as a colored metal[24, 25]. Specifically, our density functional theory (DFT) calculations show that $CaN_2$ is a nonmagnetic n-type gapped metal with the Fermi level in the principal conduction band and a band gap (we note that we use Perdew-Becke-Ernzerhof (PBE)[26] exchange-correlation (XC) functional, which usually underestimates internal band gap energy[27], but is not expected to change the physics presented in this work, see examples in Refs.[17, 28]) of 2.84 eV between principal band edges (Fig. 1a,b). The principal valence and conduction bands are dominated by N-p orbitals. The Fermi level is located 2.84 eV above the principal conduction band minimum, indicating that the transition between occupied and unoccupied states within the conduction band cannot be neglected. The optical properties of materials show the presence of strong band-to-band transitions even at low frequencies, which is consistent with the point above, especially taking into account that the energy of the first direct transition (calculated from band structure, Fig. 1b) between occupied states in the principal valence band and unoccupied states in the conduction band is 6.01 eV (Fig. 1c,d). Due to large free carrier concentration (i.e., $5.1 \times 10^{22}$ $e$/cm$^3$ corresponding to number of electrons in the principal conduction band), $CaN_2$ has a large unscreened anisotropic plasma frequency, i.e., $\omega_p^{xx} \approx 4.6$ eV, $\omega_p^{yy} \approx 4.6$ eV, and $\omega_p^{zz} \approx 6.82$ eV, which is clearly reflected in real and imaginary parts of the dielectric function (Fig. 1c). This is not surprising, as, by textbook definition (we note that a more generalized expression is actually used in DFT[29]), the Drude contribution to $\varepsilon_2(\omega)$ and $\varepsilon_1(\omega)$ can be expressed as $\varepsilon_2(\omega) = \frac{\gamma \omega_p^2}{\omega(\omega^2 + \gamma^2)}$ and $\varepsilon_1(\omega) = 1 - \frac{\omega_p^2}{(\omega^2 + \gamma^2)}$, respectively, where $\gamma$ is the damping coefficient. These expressions explicitly suggest that the Drude contribution has a substantial effect on the low-frequency region, which approaches the asymptotic value as frequency increases. This is shown in supplementary materials for all compounds discussed in this paper (Figs. S1-S4). The results above clearly demonstrate that not all gapped metals are transparent conductors, and high carrier concentrations of free carriers, while much desired for high electronic conductivity, can reduce material transparency. While this type of gapped metal is not attractive for transparent conductor applications, it has distinct features for other optoelectronic applications, including, for instance, crossing the real part of the dielectric function and strong plasmonic contribution. We also note that, in general, the properties of such materials can still be tuned if there is a way to control the free carrier concentration or effective mass by an external knob. For instance, we previously showed that the properties of gapped metals, in general, can be controlled via controllable spontaneous off-stoichiometry[15, 28, 30] or strain [20].

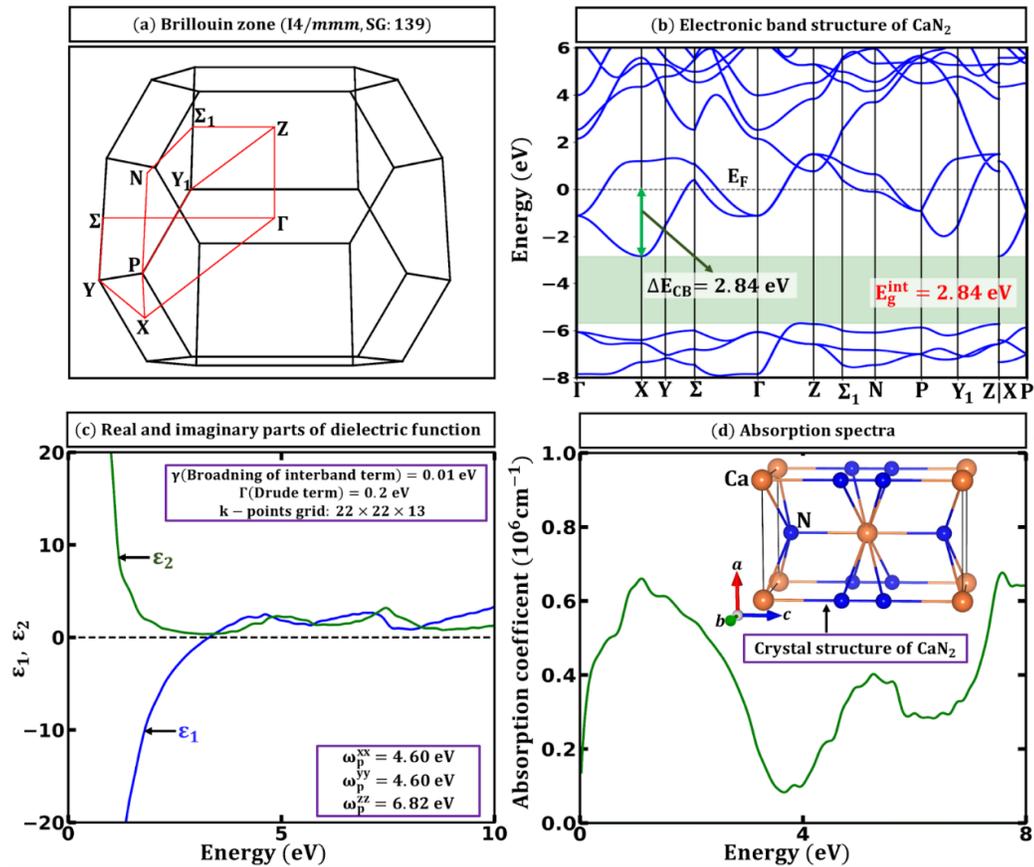

**Figure 1. Gapped metals with strong intraband absorption in the visible range.** (a) The bulk Brillouin zone for CaN$_2$ in tetragonal (I4/mmm, SG: 139) symmetry with several high-symmetry **k**-points Γ(0.0, 0.0, 0.0), X(0.0, 0.0, 0.5), N(0.0, 0.5, 0.0), Y(-0.18, 0.18, 0.5), Σ(-0.34, 0.34, 0.34), P(0.25, 0.25, 0.25), Z(0.5, 0.5, -0.5), Y1(0.5, 0.5, -0.18) and Σ1(0.34, 0.66, -0.34)[31]. The red lines highlight the high symmetry **k**-path and (b) the calculated electronic band structure along the high symmetry **k**-paths for CaN$_2$ using PBE functional. Here, the dotted line corresponds to the Fermi level, the internal band gap ($E_g^{int}$) is 2.84 eV, and the occupied part of conduction (ΔE$_{CB}$) is 2.84 eV. (c) Real and imaginary parts of the dielectric function and (d) absorption spectra of CaN$_2$ when both interband and intraband transitions are accounted for. The crystal structure of CaN$_2$ is shown as an inset.

In traditional semiconductors, the absorption spectra of a material are usually defined by the direct transition from the valence to the conduction band (i.e., direct band gap energy) with phonon assistant absorption playing significantly smaller role[32]. In a defect-free world, for traditional insulators, the color of the material is directly related to the band gap energy, and materials transparency simply requires a large band gap. However, as already noted above, for gapped metals, due to the finite occupation of the conduction or valence band, the interband transition between principal band edges can only be observed at energies significantly higher than the direct band gap energy. This resembles the Burstein-Moss effect[33, 34] in traditional wide-band-gap insulators that are self- or externally doped, and it further highlights the complexity of the optical properties of the gapped metals. We note, however, that interband transitions, even in a gapped metal, can define materials transparency. To demonstrate this, we consider the example of the NbCoSb half-Heusler compound (space group: F$\overline{4}$3m), which can be experimentally synthesized using, for instance, arc melting method[35, 36]. Here, for simplicity, we limit our investigation to stoichiometric NbCoSb compound, even though the degree of off-stoichiometry can be significant in this type of mateirals[36, 37]. Our DFT calculations show that NbCoSb is an example of n-type nonmagnetic gapped metal having the Fermi level in the conduction band, a large density of free carriers $1.9 \times 10^{22}$ $e/cm^3$, and an internal band gap of 0.93 eV between principal band edges (this internal band gap energy may be underestimated due to the using the PBE functional). Here, the conduction band is occupied within the 0.64 eV energy range. The first direct transition between the occupied principal valence band and unoccupied states in the principal conduction band is observed at 1.83 eV, implying a strong band-to-band transition in the visible range. The transparency of the material is also affected by the strong Drude contribution, the plasma frequency tensor is isotropic with a plasma frequency of 3.65 eV.

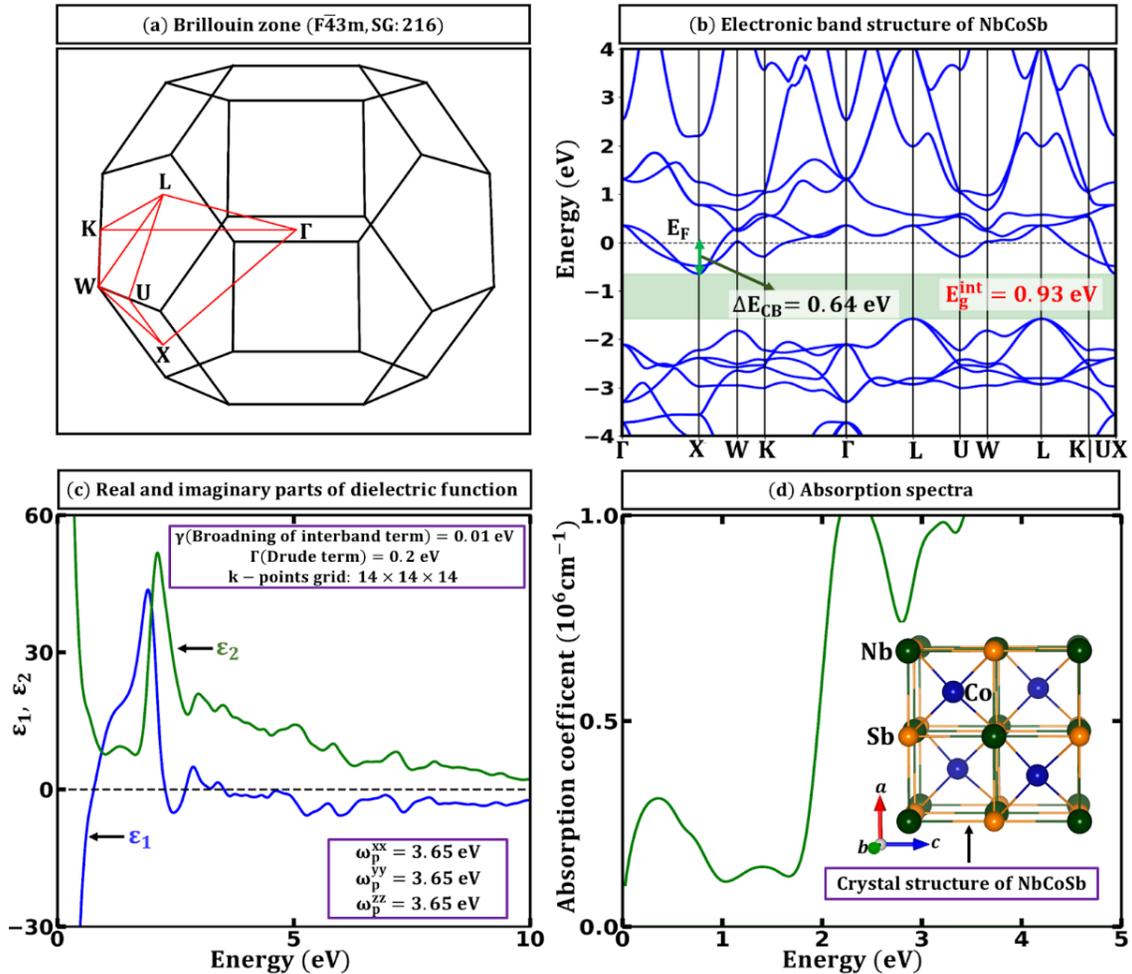

**Figure 2. Gapped metals with strong interband absorption in the visible range.** (a) The bulk Brillouin zone for NbCoSb in cubic (F$\bar{4}$3m, SG: 216) symmetry with several high-symmetry **k**-points Γ(0.0, 0.0, 0.0), X(0.5, 0.0, 0.5), W(0.5, 0.25, 0.75), K(0.375, 0.375, 0.75), U(0.625, 0.25, 0.625), and L(0.5, 0.5, 0.5)[31]. The red lines highlight the high symmetry **k**-path and (b) the calculated electronic band structure along the high symmetry **k**-paths for NbCoSb using PBE functional. Here, the dotted line corresponds to the Fermi level, the internal band gap ($E_g^{int}$) is 0.93 eV, and the occupied part of conduction (ΔE$_{CB}$) is 0.64 eV. (c) Real and imaginary parts of the dielectric function and (d) absorption spectra of NbCoSb when both interband and intraband transitions are accounted for. The crystal structure of NbCoSb is shown as an inset.

To further understand this type of compound, we consider SrNbO$_3$ gapped metal[22, 38], which is known to have distinct color experimentally (we note that experimentally this compound is often reported to be off-stoichiometric, however, for simplicity, we use the stoichiometric form here). Our calculations demonstrate that this compound has the Fermi level in the principal conduction band with a large internal band gap of 2.51 eV and free carrier concentration of 1.5×10²² $e/cm^3$. These results suggest that transitions from occupied states of the principal valance band to the principal conduction band are not activated until 3.81 eV (energy of first direct transition). Despite this, due to large free carrier concentration, there is still strong absorption in the visible range from the superposition of the transition from occupied to unoccupied states in the principal conduction band and strong Drude contribution caused by anisotropic plasma frequency contribution (i.e., $\omega_p^{xx} \approx 4.35$ eV, $\omega_p^{yy} \approx 4.39$ eV, and $\omega_p^{zz} \approx 5.02$ eV). These results thus show that having a large internal gap in gaped metal is not sufficient for making material to be a transparent conductor, and indeed transition between occupied and unoccupied states in the conduction band can noticeably reduce the materials transparency (we note, however, that both materials discussed herein in addition to interband transition also have a strong intraband transition, see supporting information).

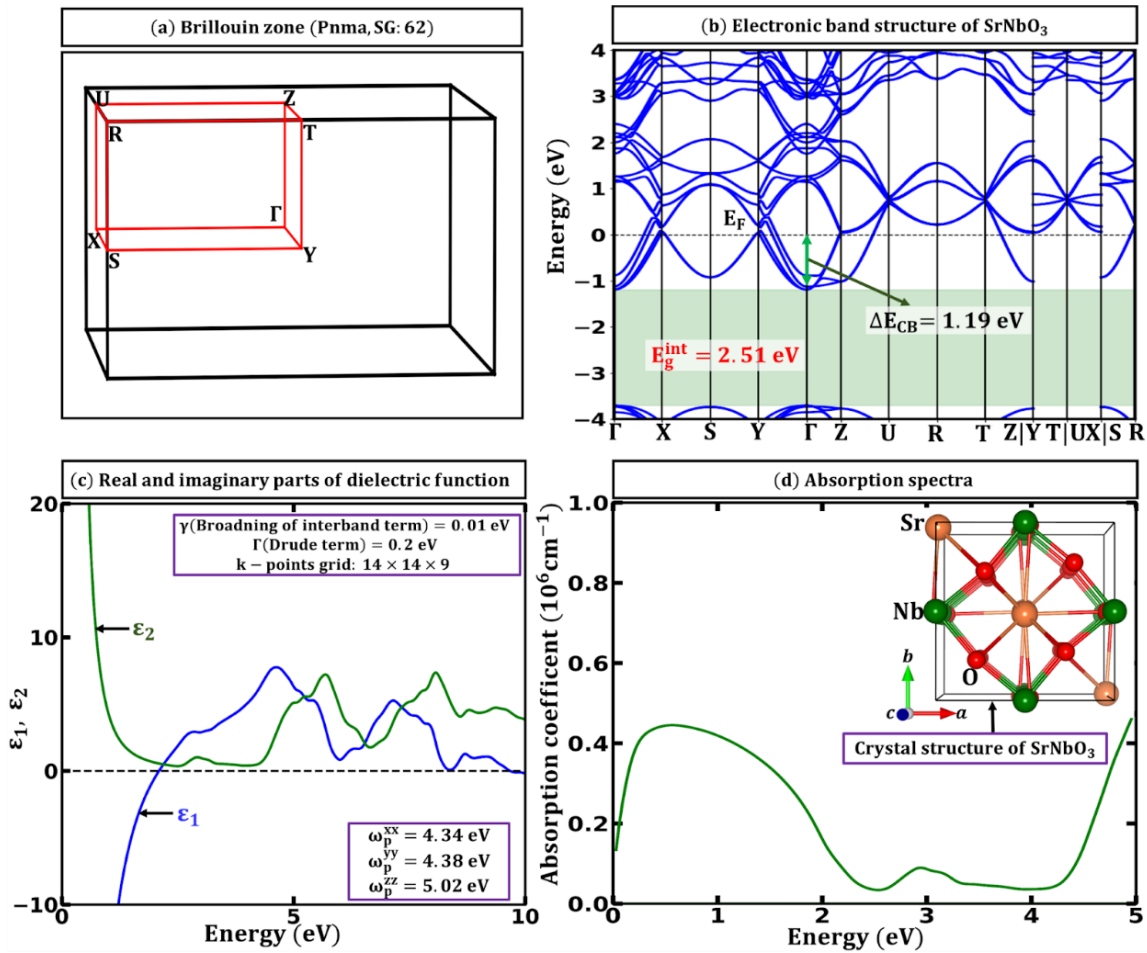

**Figure 3. Gapped metals with strong interband absorption in the visible range.** (a) The bulk Brillouin zone for SrNbO₃ in orthorhombic (Pnma, SG: 62) symmetry with several high-symmetry **k**-points Γ(0.0, 0.0, 0.0), X(0.5, 0.0, 0.0), R(0.5, 0.5, 0.5), S(0.5, 0.5, 0.0), Y(0.0, 0.5, 0.0), Z(0.0, 0.0, 0.5), U(0.5, 0.0, 0.5), R(0.5, 0.5, 0.5), and T(0.0, 0.5, 0.5)[31, 39]. The red lines highlight the high symmetry **k**-path and (b) the calculated electronic band structure along the high symmetry **k**-paths for SrNbO₃ using PBE functional. Here, the dotted line corresponds to the Fermi level, the internal band gap ($E_g^{int}$) is 0.93 eV, and the occupied part of conduction ($\Delta E_{CB}$) is 0.64 eV. (c) Average real and imaginary parts of the dielectric function and (d) absorption spectra of SrNbO₃ when both interband and intraband transitions are accounted for. The crystal structure of SrNbO₃ is shown as an inset.

The above results demonstrate that identification of the gapped metals that are both conductive and transparent is a complex task. Thus, the gapped metal as a transparent conductor should satisfy the following conditions: (i) the compound needs to be realizable, i.e., one that can be really synthesized in stoichiometric form (we note that indeed first principle literature is full of fantasy materials that cannot be realized[40, 41]); (ii) the compound should have the free carrier concentration to provide sufficient electronic transport, however, this concentration should be not too large to not cause strong Drude absorption is visible range making gapped metal to be opaque; (iii) the compound should have minimized transition from occupied to unoccupied states in the visible range. To demonstrate the example of such material, we consider the case of $Sr_5Nb_5O_{17}$ n-type gapped metal, which was previously prepared by the floating-zone melting technique and has been characterized by single-crystal X-ray analysis to have an orthorhombic Pnnm space group[42]. The principal valence band of the compound is mainly determined by O-p states, while the principal conduction band is defined by Nb-d states. Our PBE calculations show that the compound is a nonmagnetic n-type gapped metal that has 1e per f.u. (i.e., $1.3 \times 10^{21}$ $e$/cm$^3$) in the principal conduction band, which is spread in 0.63 eV energy space, and the large internal band gap energy of 2.3 eV. The first available direct transition from occupied states of principal valence to unoccupied states of principial conduction band is 3.12 eV, resulting in band-to-band transition from the principal valence to the principal conduction band within the visible range (Fig. 4). The free carrier concentration is large enough to provide sufficiently high electronic conductivity but is low enough not to have strong free carrier absorption in the visible range. For instance, the anisotropic plasma frequency tensor (i.e., $\omega_p^{xx} \approx 1.95$ eV, $\omega_p^{yy} \approx 0.49$ eV, and $\omega_p^{zz} \approx 0.11$ eV) do not results is substantial free carrier absorption. Hence, the resulted light absorption in the visible range is limited, making $Sr_5Nb_5O_{17}$ a potential transparent conductor.

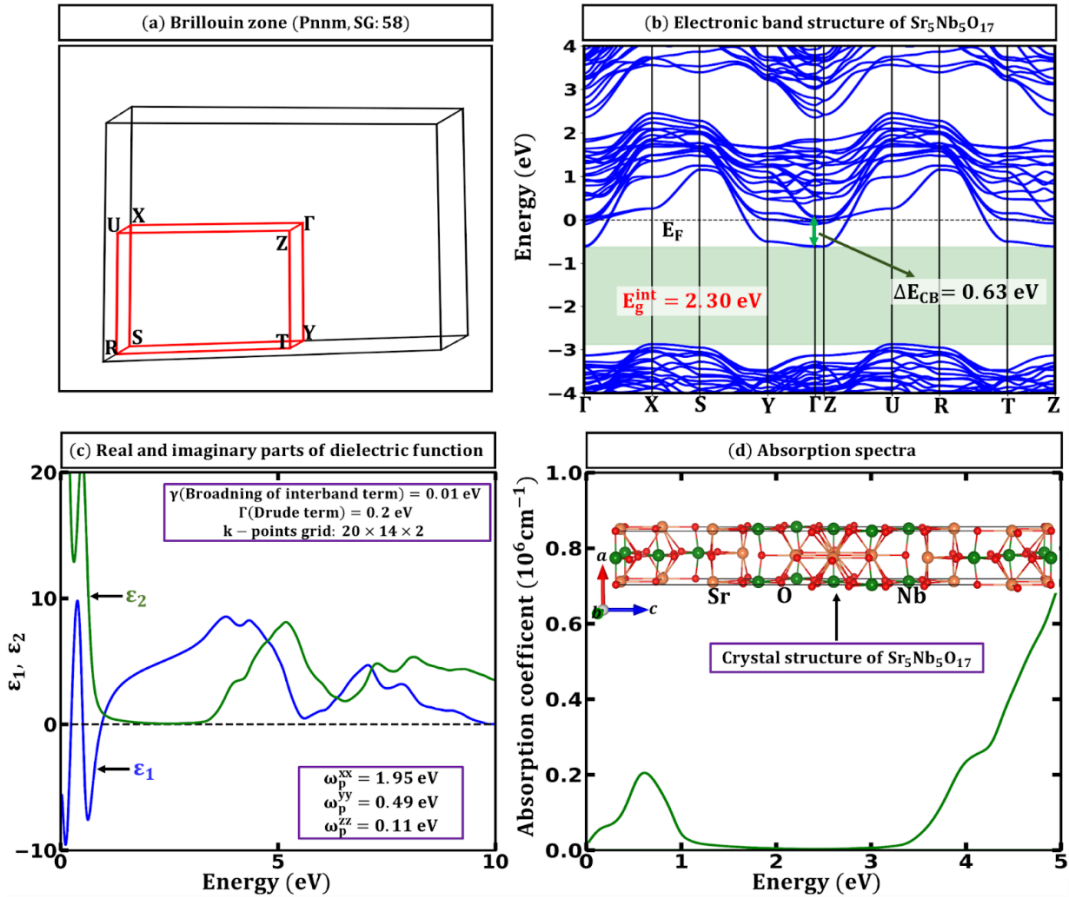

**Figure 4. Gapped metals that are potential transparent conductors.** (a) The bulk Brillouin zone for $Sr_5Nb_5O_{17}$ in orthorhombic (Pnnm, SG: 58) symmetry with several high-symmetry **k**-points Γ(0.0, 0.0, 0.0), X(0.5, 0.0, 0.0), S(0.5, 0.5, 0.0), Y(0.0, 0.5, 0.0), Z(0.0, 0.0, 0.5), U(0.5, 0.0, 0.5), R(0.5, 0.5, 0.5), and T(0.0, 0.5, 0.5)[39]. The red lines highlight the high symmetry **k**-path and (b) the calculated electronic band structure along the high symmetry **k**-paths for $Sr_5Nb_5O_{17}$ using PBE functional. Here, the dotted line corresponds to the Fermi level, the internal band gap ($E_g^{int}$) is 2.30 eV, and the occupied part of conduction ($\Delta E_{CB}$) is 0.63 eV. (c) Average real and imaginary parts of the dielectric function and (d) absorption spectra of $Sr_5Nb_5O_{17}$ when both interband and intraband transitions are accounted for. The crystal structure of $Sr_5Nb_5O_{17}$ is shown as an inset.

In summary, we have investigated the optical properties of gapped metals in the context of their potential use as transparent conductors, demonstrating their correlation to carrier concentration, effective mass, and the internal band gap. Our study focuses on three types of gapped metals: (i) those with strong intraband absorption in the visible range, (ii) those with strong interband absorption in the visible range, and (iii) those with optimal carrier concentration and optimal internal band gap. For the first category, we analyzed $CaN_2$ and found that large free carrier concentration results in strong band-to-band transitions and large unscreened plasma frequency, reducing materials transparency. These results demonstrate that a large internal gap in gapped metals is insufficient for achieving transparency and that this type of materials is not attractive for transparent conductor applications but may have other optoelectronic applications. The second category, represented by NbCoSb and $SrNbO_3$, also shows strong absorption in the visible range due to the superposition of strong interband transitions or large free carrier concentrations. This type of materials highlights that a small internal gap can or intraband transition in the principal conduction band can make materials unsuitable for transparent conductor application. Lastly, we considered $Sr_5Nb_5O_{17}$, which has optimal carrier concentration and internal band gap. This compound satisfies the conditions required for transparent conductors.

The authors thank the "ENSEMBLE[3] - Centre of Excellence for nanophotonics, advanced materials and novel crystal growth-based technologies" project (GA No. MAB/2020/14) carried out within the International Research Agendas programme of the Foundation for Polish Science co-financed by the European Union under the European Regional Development Fund and the European Union's Horizon 2020 research and innovation programme Teaming for Excellence (GA. No. 857543) for support of this work. We gratefully acknowledge Poland's high-performance computing infrastructure PLGrid (HPC Centers: ACK Cyfronet AGH) for providing computer facilities and support within computational grant no. PLG/2022/015458.